\documentclass[10pt,twocolumn]{IEEEtran}
\usepackage[latin1]{inputenc}
\usepackage{amssymb}
\usepackage{latexsym}
\usepackage{mathrsfs}
\usepackage{amsfonts}
\usepackage{amsbsy}
\usepackage{amsmath}
\usepackage{times}
\usepackage{epsfig,epsf,times,amssymb,amsmath}
\usepackage{color}
\usepackage{setspace}
\usepackage{float}
\usepackage{subfigure}
\usepackage{graphicx}

\title{Adaptive Delay-Tolerant Distributed Space-Time
Coding Based on Adjustable Code Matrices for Cooperative MIMO Relaying Systems}

\author{\IEEEauthorblockN{Tong Peng, Rodrigo C. de Lamare}\\
\IEEEauthorblockA{Communications Reasearch Group, Department of Electronics\\University of York, York YO10 5DD, UK\\
Email: tp525@ohm.york.ac.uk; rcdl500@ohm.york.ac.uk}}

\begin{document}

\maketitle

\IEEEpeerreviewmaketitle

\begin{abstract}
An adaptive delay-tolerant distributed space-time coding (DSTC)
scheme that exploits feedback is proposed for two-hop cooperative
MIMO networks. Maximum likelihood (ML) receivers and adjustable code
matrices are considered subject to a power constraint with a
decode-and-forward (DF) cooperation strategy. In the proposed
delay-tolerant DSTC scheme, an adjustable code matrix is employed to
transform the space-time coded matrix at the relay nodes.
Least-squares (LS) algorithms are developed with reduced
computational complexity to adjust the parameters of the codes.
Simulation results show that the proposed algorithms obtain
significant performance gains and address the delay issue for
cooperative MIMO systems as compared to existing delay-tolerant DSTC
schemes.
\end{abstract}


\section{Introduction}

Cooperative multiple-input and multiple-output (MIMO) systems can
obtain diversity gains by providing copies of the transmitted
signals with the help of relays to improve the reliability of
wireless communication systems \cite{J.N.Laneman2004}-\cite{tds2}.
The basic idea behind these cooperative relaying  systems is to
employ multiple relay nodes between the source node and the
destination node to form a distributed antenna array, which can
provide significant advantages in terms of diversity gains. Several
cooperation strategies that exploit the links between the relay
nodes and the destination node such as amplify-and-forward (AF),
decode-and-forward (DF), compress-and-forward (CF)
\cite{J.N.Laneman2004} and distributed space-time coding (DSTC)
schemes \cite{J.N.Laneman2003}, \cite{Yiu S.} that employ space-time
codes \cite{Yiu S.,RC De Lamare} have been extensively studied in
the literature. A key problem that arises in the cooperative MIMO
systems and which degrades the performance such systems is the
existence of delays between the signals that are space-time coded at
the relays and decoded at the destination.

Delay-tolerant Space-Time Coding (DT-STC) schemes
\cite{Gamal}-\cite{Bhatnagar} which address the delay between the
relay nodes have been recently considered for distributed MIMO
systems. Extending the distributed threaded algebraic space-time
(TAST) codes \cite{Gamal}, Damen and Hammons designed a
delay-tolerant coding scheme in \cite{Damen} by the extension of the
Galois group and field employed in the coding scheme to achieve full
diversity and full rate. A further optimization which ensures that
the codes in \cite{Damen} have full diversity with the minimum
length and lower decoding complexity is presented in
\cite{Torbatian}. The authors of \cite{Zhimeng} propose
delay-tolerant distributed linear convolutional STCs which can
maintain the full diversity property under any delay profile. In
\cite{Bhatnagar}, a precoder based on the linear constellation
precoder (LCP) design is employed to construct an optimal DT-STC to
achieve the upper bound of the error probability.

In the literature so far, two basic configurations of distributed
space-time schemes have been reported: one in which the coding is
performed independently at the relays \cite{Birsen
Sirkeci-Mergen,TARMO} and another in which coding is performed
across the relays \cite{Damen,Bhatnagar}. It is not clear the key
advantages of these schemes and their suitability for situations
with delays. In addition, the work on delay-tolerant space-time
codes has focused on schemes that do not employ feedback to improve
the performance of the systems. Strategies to exploit feedback and
improve the design of distributed space-time coding schemes remain
unexplored.

In this paper, we propose an adaptive delay-tolerant distributed
space-time coding scheme that employs feedback and devise a design
algorithm for cooperative MIMO relaying systems with delays at the
relay nodes. A delay-tolerant adjustable code matrices optimization
(DT-ACMO) algorithm based on the the ML criterion subject to
constraints on the transmitted power at the relays for different
cooperative systems is presented. Adaptive optimization algorithms
using least-squares (LS) estimation method are developed for the
DT-ACMO scheme in order to release the destination node from the
high computational complexity of the optimization process. In
particular, we consider network configurations in which the coding
is performed across the relays and independently at the relays to
study existing and the proposed delay-tolerant distributed
space-time coding scheme. We study how the adjustable code matrix
affects the DSTBC during the encoding process and how to optimize
the adjustable code matrices by employing an ML detector.

The paper is organized as follows. Section II introduces two
different types of two-hop cooperative MIMO systems with multiple
relays applying the DF strategy and DSTC schemes. In Section III the
proposed optimization algorithm for the adjustable code matrix is
derived, and the results of the simulations are given in Section IV.
Section V gives the conclusions of the work.

\section{Cooperative MIMO System Model}

\begin{figure}[htb]
\label{Fig.1}
\begin{center}
  \includegraphics[width=3.5in]{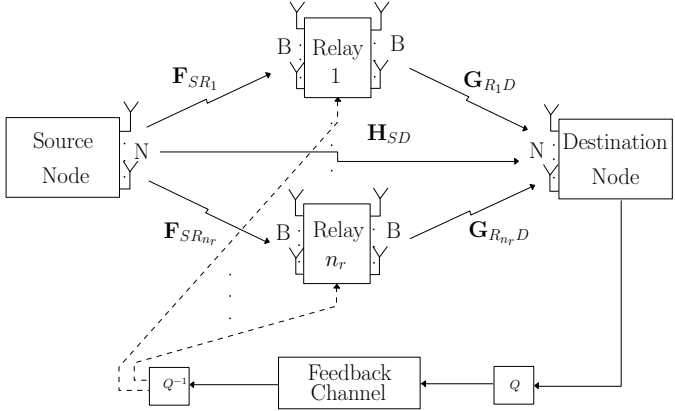}
  \caption{Cooperative MIMO system model with an arbitrary number of antennas at the relays
  and $n_r$ relay.
nodes}\label{1}
\end{center}
\end{figure}

We consider a cooperative communication system, which consists of
one source node, $n_r$ relay nodes (Relay 1, Relay 2, ..., Relay
$n_r$), and one destination node as shown in Fig. 1. A two-hop
cooperative MIMO system with all nodes employ $N$ antennas and each
of them can either transmit or receive at one time. We consider two
types of system configurations in this paper, one is called the
multiple antenna systems (MAS) configuration which employs relay
nodes with $B=N$ antennas, and another one is called the single
antenna systems (SAS) configuration which employs relay nodes with
$B=1$ antenna. A delay profile ${\boldsymbol
\Delta}=[\delta_1,\delta_2, ..., \delta_{n_r}]$, where $\delta_k$
denotes the relative delay of the signal received from the $k$th
relay node as a reference to the earliest received relay signal. Let
${\boldsymbol s}[i]$ denote the transmitted information symbol
vector at the source node, which is derived as ${\boldsymbol s}[i] =
[s_{1}[i], s_{2}[i], ... , s_{N}[i]]$, and has a covariance matrix
$E\big[ {\boldsymbol s}[i]{\boldsymbol s}^{H}[i]\big] =
\sigma_{s}^{2}{\boldsymbol I}_N$, where $\sigma_s^2$ is the signal
power which we assume to be equal to 1. The source node broadcasts
${\boldsymbol s}[i]$ from the source to $n_r$ relay nodes as well as
to the destination node in the first hop, and all the relay nodes
can detect the received symbol vectors with no errors.

Since the DF protocol is employed at the relay nodes, the received symbols are detected and re-encoded at each relay node prior to transmission to the destination in the second hop. We first consider the $k$th relay node in MAS, the $N \times 1$ signal vector ${\boldsymbol s}[i]$ will be re-encoded by an $N \times T$ DSTC scheme ${\boldsymbol M}({\boldsymbol s})$, multiplied by an $N \times N$ adjustable code matrix ${\boldsymbol \Phi}_k[i]$ generated randomly \cite{Birsen Sirkeci-Mergen}, and then forwarded to the destination node. The relationship between the $k$th relay and the destination node can be described as
\begin{equation}\label{2.1}
\begin{aligned}
{\boldsymbol R}_{R_{k}D_{MAS}}[i] = & {\boldsymbol G}^{\boldsymbol \Delta}_{{R_kD}_{MAS}}[i]{\boldsymbol \Phi}_{k_{MAS}}[i]{\boldsymbol M}_{R_{k}D_{MAS}}[i]\\ & + {\boldsymbol N}_{R_{k}D}[i],
\end{aligned}
\end{equation}
where ${\boldsymbol G}^{\boldsymbol \Delta}_{R_kD_{MAS}}[i]$ denotes the $(\delta_{max}+N) \times N$ delayed version of the channel matrix between the $k$th relay node and the destination node, and ${\boldsymbol M}_{R_{k}D_{MAS}}[i]$ is the $N \times T$ DSTC scheme. ${\boldsymbol N}_{R_{k}D}[i]$ stands for the noise matrix at the destination node with variance $\sigma^2_d$. The $(\delta_{max} + N) \times T$ received symbol matrix ${\boldsymbol R}_{R_{k}D_{MAS}}[i]$ in (\ref{2.1}) can be written as a $(\delta_{max} + NT) \times 1$ vector ${\boldsymbol r}_{R_kD_{MAS}}[i]$ given by
\begin{equation}\label{2.2}
{\boldsymbol r}_{R_kD_{MAS}}[i]  = {\boldsymbol \Delta}_k{\boldsymbol \Phi}_{{eq_k}_{MAS}}[i]{\boldsymbol
G}_{{eq_k}_{MAS}}[i]{\boldsymbol s}[i] + {\boldsymbol n}_{R_{k}D}[i],
\end{equation}
where the $N(\delta_{max}+T) \times NT$ delay profile matrix ${\boldsymbol \Delta}_k = [{\boldsymbol 0}_{\delta_k \times N} ; {\boldsymbol I}_N ; {\boldsymbol 0}_{(\delta_{max}-\delta_k) \times N}]$ at the $k$th relay nodes is considered. The block diagonal $NT \times NT$ matrix ${\boldsymbol \Phi}_{{eq_k}_{MAS}}[i]$ denotes the equivalent adjustable code matrix and the $NT \times N$ matrix ${\boldsymbol G}_{{eq_k}_{MAS}}[i]$ stands for the equivalent channel matrix which is the DSTC scheme ${\boldsymbol M}({\boldsymbol s})$ combined with the channel matrix ${\boldsymbol G}_{R_kD_{MAS}}[i]$. The $N(\delta_{max} + T) \times 1$ equivalent noise vector ${\boldsymbol n}_{R_{k}D}[i]$ generated at the destination node contains the noise parameters.

In SAS, only $1$ antenna is employed at each relay node and the $n_r \times T$ DSTC scheme ${\boldsymbol M}({\boldsymbol s})$ is encoded across all the relay nodes. At the $k$th relay node, the $1 \times T$ code vector ${\boldsymbol m}_k({\boldsymbol s})$ denotes the $k$th row in ${\boldsymbol M}({\boldsymbol s})$ which is allocated. The received signal matrix from the $k$th relay node at the destination node can be described as
\begin{equation}\label{2.1-1}
\begin{aligned}
{\boldsymbol R}_{R_kD_{SAS}}[i] = & {\boldsymbol g}^{\boldsymbol \Delta}_{R_kD_{SAS}}[i]({\boldsymbol \phi}_{k_{SAS}}[i]\cdot{\boldsymbol m}_{R_kD_{SAS}}[i]) \\ & + {\boldsymbol N}_{R_{k}D}[i],
\end{aligned}
\end{equation}
where ${\boldsymbol g}^{\boldsymbol \Delta}_{R_kD_{SAS}}[i]$ denotes the $(\delta_{max}+N) \times 1$ delayed version of the channel vector between the $k$th relay node and the destination node, and ${\boldsymbol \phi}_k[i]$ is the $1 \times T$ adjustable code vector. ${\boldsymbol N}_{R_{k}D}[i]$ stands for the noise matrix with variance $\sigma^2_d$. The $(\delta_{max} + N) \times T$ received symbol matrix ${\boldsymbol R}_{R_kD_{SAS}}[i]$ in (\ref{2.1-1}) can be written as a $(\delta_{max} + NT) \times 1$ vector ${\boldsymbol r}_{R_kD_{SAS}}[i]$ given by
\begin{equation}\label{2.2-1}
{\boldsymbol r}_{R_kD_{SAS}}[i]  = {\boldsymbol \Delta}_k{\boldsymbol G}_{{eq_k}_{SAS}}[i]{\boldsymbol \Phi}_{{eq_k}_{SAS}}[i]{\boldsymbol s}[i] + {\boldsymbol n}_{R_{k}D}[i],
\end{equation}
where the block diagonal $N \times N$ matrix ${\boldsymbol \Phi}_{{eq_k}_{SAS}}[i]$ denotes the diagonal equivalent adjustable code matrix and the block diagonal $NT \times N$ matrix ${\boldsymbol G}_{{eq_k}_{SAS}}[i]$ stands for the equivalent channel matrix. The $N(\delta_{max} + T) \times 1$ equivalent noise vector ${\boldsymbol n}_{R_kD}[i]$ generated at the destination node contains the noise parameters.

After rewriting the received vectors in (\ref{2.2}) and (\ref{2.2-1}) we can consider the
received symbol vector at the destination node as a $(T+1)N \times 1$
vector. Therefore, the received symbol vector for MAS and SAS can be written as
\begin{equation}\label{2.3}
\begin{aligned}
{\boldsymbol r}_{MAS}[i]  & = {\boldsymbol D}_{D_{MAS}}[i]{\boldsymbol s}[i]  + {\boldsymbol n}_D[i],\\
{\boldsymbol r}_{SAS}[i]  & = {\boldsymbol D}_{D_{SAS}}[i]{\boldsymbol s}[i] + {\boldsymbol n}_D[i],
\end{aligned}
\end{equation}
where the $N(\delta_{max} + T + 1) \times N$ block diagonal matrices ${\boldsymbol D}_D[i]$ in MAS and SAS denotes the channel gain matrix of all the links in the network. We assume that the coefficients in all channel matrices are independent and remain constant over the transmission. The $N(\delta_{max} + T + 1) \times N$ noise vector ${\boldsymbol n}_D[i]$ contains the equivalent received noise vector at the destination node, which can be modeled as complex Gaussian random cariables with zero mean and covariance $\sigma^{2}_d$.

\section{Delay-Tolerant Adjustable Code Matrix Optimization for Delayed DSTCs}

In this section, we propose the design of a delay-tolerant adjustable code matrix optimization (DT-ACMO) strategy with ML receivers for DSTC schemes in MAS and SAS. An adaptive recursive least square (RLS) algorithm \cite{S.Haykin} for determining the adjustable code matrices with reduced complexity is devised. The adjustable code matrix is computed at the destination node and obtained by a feedback channel. It is worth to mention that the code matrices are only used at the relay node so the direct link from the source node to the destination node is not considered here for simplicity.

\subsection{DT-ACMO Algorithm for MAS}

We employ the ML criterion to determine the code matrices for MAS, then we have to store an $N \times D$ matrix ${\boldsymbol S}$ at the destination node which contains all the possible combinations of the transmitted symbol vectors. The constrained ML optimization problem can be written as
\begin{equation}\label{4.1}
\begin{aligned}
     & [\hat {\boldsymbol s}[i], \hat {\boldsymbol \Phi}_{{eq_k}_{MAS}}[i]] \\ & = \arg\min_{{\boldsymbol s}[i],{\boldsymbol \Phi}_{{eq_k}_{MAS}}[i]} \|{\boldsymbol r}_{MAS}[i] - \hat {\boldsymbol r}_{MAS}[i]\|^2, \\
     & ~~~s.t.~ \sum_{k=1}^{n_r}\rm{Tr}({\boldsymbol \Phi}_{{eq_k}_{MAS}}[i]{\boldsymbol \Phi}^\emph{H}_{{eq_k}_{MAS}}[i])\leq \rm{P_R},
\end{aligned}
\end{equation}
where $\hat {\boldsymbol r}_{MAS}[i]= \sum_{k=1}^{n_r}{\boldsymbol \Phi}^{\boldsymbol \Delta}_{{eq_k}_{MAS}}[i]{\boldsymbol G}_{{eq_k}_{MAS}}[i]\hat{\boldsymbol s}[i]$ denotes the received symbol vector without noise which is determined by substituting each column of ${\boldsymbol S}$ into (\ref{4.1}), and ${\boldsymbol \Phi}^{\boldsymbol \Delta}_{{eq_k}_{MAS}}[i]$ denotes the $N(\delta_{max} + T) \times NT$ delayed adjustable code matrix at the $k$th relay node. As mentioned in \cite{TARMO}, the optimization algorithm contains a discrete part which refers to the ML detection and a continuous part which refers to the optimization of the adjustable code matrix. The detection and the optimization can be implemented separately as they do not depend on each other. As a result, other detectors such as MMSE and sphere decoders can be used in the detection part in order to reduce the computational complexity. The Lagrangian expression of the optimization problem in (\ref{4.1}) is given by
\begin{equation}\label{4.2}
\begin{aligned}
    \mathscr{L} = & \|{\boldsymbol r}_{MAS}[i]-\sum_{k=1}^{n_r}{\boldsymbol \Phi}^{\boldsymbol \Delta}_{{eq_k}_{MAS}}[i]{\boldsymbol G}_{{eq_k}_{MAS}}[i]\hat{\boldsymbol s}[i]\|^2 \\ & +\lambda(\sum_{k=1}^{n_r}Tr({\boldsymbol \Phi}_{{eq_k}_{MAS}}[i]{\boldsymbol \Phi}^\emph{H}_{{eq_k}_{MAS}}[i])-P_R).
\end{aligned}
\end{equation}
After determining the transmitted symbol vector $\hat {\boldsymbol s}[i]$ by ML detection, we can calculate the optimal adjustable code matrix ${\boldsymbol \Phi}^{\boldsymbol \Delta}_{{eq_k}_{MAS}}[i]$ by employing the RLS estimation algorithm. The adjustable code matrix ${\boldsymbol \Phi}^{\boldsymbol \Delta}_{{eq_k}_{MAS}}[i]$ will be updated after a number of iterations and finally achieve its optimum by the RLS estimation algorithm with a reduced computational complexity. The superior convergence behavior of LS type algorithms when the size of the adjustable code matrix is large is the reason for its utilization. It is worth to mention that the computational complexity reduces from cubic to square by employing the RLS algorithm.

In order to derive an RLS algorithm, the optimization problem is given by
\begin{equation}\label{4.5}
\begin{aligned}
    & [\hat {\boldsymbol \Phi}^{\boldsymbol \Delta}_{{eq_k}_{MAS}}[i]] = \arg\min_{{\boldsymbol \Phi}^{\boldsymbol \Delta}_{{eq_k}_{MAS}}[i]} \sum_{n=1}^{i}\lambda^{i-n}\|{\boldsymbol r}[n]-{\boldsymbol {\hat{r}}}[i]\|^2, \\
    & ~~~s.t.~ \sum_{k=1}^{n_r}\rm{Tr}({\boldsymbol \Phi}^{\boldsymbol \Delta}_{{eq_k}_{MAS}}[i]({\boldsymbol \Phi}^{\boldsymbol \Delta}_{{eq_k}_{MAS}}[i])^\emph{H}[i])\leq \rm{P_R},
\end{aligned}
\end{equation}
where $\lambda$ stands for the forgetting factor. By expanding the right-hand side of (\ref{4.5}) and taking the gradient with respect to $({\boldsymbol \Phi}^{\boldsymbol \Delta}_{{eq_k}_{MAS}}[i])^*$ and equating the terms to zero, we obtain
\begin{equation}\label{4.6}
\begin{aligned}
    {\boldsymbol \Phi}^{\boldsymbol\Delta}_{{eq_k}_{MAS}}[i] = & (\sum_{n=1}^{i}\lambda^{i-n}{\boldsymbol r}_{e_{MAS}}[n]{\boldsymbol r}^\emph{H}_{k_{MAS}}[n]) \\ & (\sum_{i=1}^{n}\lambda^{i-n}{\boldsymbol r}_{k_{MAS}}[n]{\boldsymbol r}^\emph{H}_{k_{MAS}}[n])^{-1},
\end{aligned}
\end{equation}
where the $N(\lambda_{max}+T) \times 1$ vector ${\boldsymbol r}_{e_{MAS}}[n]={\boldsymbol \Delta}_k{\boldsymbol \Phi}_{{eq_k}_{MAS}}[i]{\boldsymbol G}_{{eq_k}_{MAS}}[i]{\boldsymbol s}[i]$ and ${\boldsymbol r}_{k_{MAS}}[n]={\boldsymbol G}_{{eq_k}_{MAS}}[i]{\boldsymbol s}[i]$. The power constraint is still not considered during the derivation but is enforced by the normalization strategy after the optimization. We can define
\begin{equation}\label{4.7}
\begin{aligned}
    {\boldsymbol \Psi}[i]  & = \sum_{n=1}^{i}\lambda^{i-n}{\boldsymbol r}_{k_{MAS}}[n]{\boldsymbol r}^\emph{H}_{k_{MAS}}[n] \\ & = \lambda {\boldsymbol \Psi}[i-1]+{\boldsymbol r}_{k_{MAS}}[n]{\boldsymbol r}^\emph{H}_{k_{MAS}}[n],
\end{aligned}
\end{equation}
\begin{equation}\label{4.8}
\begin{aligned}
    {\boldsymbol Z}[i] & = \sum_{n=1}^{i}\lambda^{i-n}{\boldsymbol r}_{e_{MAS}}[n]{\boldsymbol r}^\emph{H}_{k_{MAS}}[n] \\ & =\lambda {\boldsymbol Z}[i-1]+{\boldsymbol r}_{e_{MAS}}[n]{\boldsymbol r}^\emph{H}_{k_{MAS}}[n],
\end{aligned}
\end{equation}
so that we can rewrite (\ref{4.6}) as
\begin{equation}\label{4.9}
    {\boldsymbol \Phi}^{\boldsymbol\Delta}_{{eq_k}_{MAS}}[i] = {\boldsymbol Z}[i]{\boldsymbol \Psi}^{-1}[i].
\end{equation}
By employing the matrix inversion lemma in \cite{Tylavsky}, we can obtain
\begin{equation}\label{4.10}
    {\boldsymbol \Psi}^{-1}[i]=\lambda^{-1}{\boldsymbol \Psi}^{-1}[i-1]-\lambda^{-1}{\boldsymbol k}[i]{\boldsymbol r}^\emph{H}_{k_{MAS}}[i]{\boldsymbol \Psi}^{-1}[i-1],
\end{equation}
where ${\boldsymbol k}[i]=(\lambda^{-1}{\boldsymbol \Psi}^{-1}[i-1]{\boldsymbol r}_{k_{MAS}}[i])/(1+\lambda^{-1}{\boldsymbol r}^\emph{H}_{k_{MAS}}[i]{\boldsymbol \Psi}^{-1}[i-1]{\boldsymbol r}_{k_{MAS}}[i])$. We define ${\boldsymbol P}[i]={\boldsymbol \Psi}^{-1}[i]$ and by substituting (\ref{4.8}) and (\ref{4.10}) into (\ref{4.9}), the expression of the code matrix is given by
\begin{equation}\label{4.11}
\begin{aligned}
    & {\boldsymbol \Phi}^{\boldsymbol\Delta}_{{eq_k}_{MAS}}[i]\\
    &= \lambda{\boldsymbol Z}[i-1]{\boldsymbol P}[i]+{\boldsymbol r}_{e_{MAS}}[i]{\boldsymbol r}^\emph{H}_{k_{MAS}}[i]{\boldsymbol P}[i]\\ &= {\boldsymbol Z}[i-1]{\boldsymbol P}[i-1]+{\boldsymbol Z}[i-1]{\boldsymbol k}[i]{\boldsymbol r}^\emph{H}_{k_{MAS}}[i]{\boldsymbol P}[i-1]\\ & ~~~+{\boldsymbol r}_{e_{MAS}}[i]{\boldsymbol r}^\emph{H}_{k_{MAS}}[i]{\boldsymbol P}[i]\\ &= {\boldsymbol \Phi}^{\boldsymbol\Delta}_{{eq_k}_{MAS}}[i-1]\\ & ~~~ +\lambda^{-1}({\boldsymbol r}_{e_{MAS}}[i]-{\boldsymbol Z}[i-1]{\boldsymbol k}[i]){\boldsymbol r}^\emph{H}_{k_{MAS}}[i]{\boldsymbol P}[i-1].
\end{aligned}
\end{equation}
As mentioned before, a normalization process after (\ref{4.11}) is implemented as ${\boldsymbol \Phi}^{\boldsymbol\Delta}_{{eq_k}_{MAS}}[i] = \frac{\sqrt{\rm{P_R}}{\boldsymbol \Phi}^{\boldsymbol\Delta}_{{eq_k}_{MAS}}[i]}{\sqrt{\sum_{k=1}^{n_r}\rm{Tr}({\boldsymbol \Phi}^{\boldsymbol\Delta}_{{eq_k}_{MAS}}[i]({\boldsymbol \Phi}^{\boldsymbol\Delta}_{{eq_k}_{MAS}}[i])^\emph{H})}}$ in order to maintain the energy of the code matrices. The   Table I shows a summary of the DT-ACMO RLS algorithm.
\begin{table}
  \caption{Summary of the DT-ACMO RLS Algorithm}\label{}
  \begin{tabular}{cc}
  \hline
1: & Initialize: ${\boldsymbol P}[0]=\delta^{-1}{\boldsymbol I}_{N(\lambda_{max}+T) \times N(\lambda_{max}+T)}$, \\
2: & ${\boldsymbol Z}[0] = {\boldsymbol I}_{N(\lambda_{max}+T) \times N(\lambda_{max}+T)}$, \\
3: & generate ${\boldsymbol \Phi}[0]$ randomly \\
4: & with the power constraint. \\
5: & For each instant of time, $i$=1, 2, ..., compute \\
6: & ${\boldsymbol k}[i]=\frac{\lambda^{-1}{\boldsymbol \Psi}^{-1}[i-1]{\boldsymbol r}_{k_{MAS}}[i]}{1+\lambda^{-1}{\boldsymbol r}_{k_{MAS}}^\emph{H}[i]{\boldsymbol \Psi}^{-1}[i-1]{\boldsymbol r}_{k_{MAS}}[i]}$,\\
7: & update ${\boldsymbol \Phi}^{\boldsymbol\Delta}_{{eq_k}_{MAS}}[i]$ by (\ref{4.11}), \\
8: & ${\boldsymbol P}[i]=\lambda^{-1}{\boldsymbol P}[i-1]-\lambda^{-1}{\boldsymbol k}[i]{\boldsymbol r}_{k_{MAS}}^\emph{H}[i]{\boldsymbol P}[i-1]$, \\
9: & ${\boldsymbol Z}[i] = \lambda {\boldsymbol Z}[i-1]+{\boldsymbol r}_{e_{MAS}}[i]{\boldsymbol r}_{k_{MAS}}^\emph{H}[i]$. \\
10: & ${\boldsymbol \Phi}^{\boldsymbol\Delta}_{{eq_k}_{MAS}}[i] = \frac{\sqrt{\rm{P_R}}{\boldsymbol \Phi}^{\boldsymbol\Delta}_{{eq_k}_{MAS}}[i]}{\sqrt{\sum_{k=1}^{n_r}\rm{Tr}({\boldsymbol \Phi}^{\boldsymbol\Delta}_{{eq_k}_{MAS}}[i]({\boldsymbol \Phi}^{\boldsymbol\Delta}_{{eq_k}_{MAS}}[i])^\emph{H})}}$.\\
\hline
\end{tabular}
\vspace{-2em}
\end{table}

\subsection{DT-ACMO Algorithm for SAS}

In SAS model, the main difference of the proposed algorithm is the commotional complexity which is related to the dimension of the adjustable code matrix. We only consider the received symbols from the relay nodes so that the equivalent received vector at the destination node is expressed as
\begin{equation}\label{4.13}
    {\boldsymbol r}_{SAS}[i]=\sum_{k=1}^{n_r}{\boldsymbol \Delta}_k{\boldsymbol G}_{{eq_k}_{SAS}}[i]{\boldsymbol \Phi}_{{eq_k}_{SAS}}[i]{\boldsymbol s}[i] + {\boldsymbol n}_{RD}[i],
\end{equation}
According to the ML criteria, the optimization problem can be written as
\begin{equation}\label{4.14}
\begin{aligned}
     & [\hat {\boldsymbol s}[i], \hat {\boldsymbol \Phi}_{{eq_k}_{SAS}}[i]] \\
& = \arg\min_{{\boldsymbol s}[i],{\boldsymbol \Phi}_{{eq_k}_{SAS}}[i]} \|{\boldsymbol r}_{SAS}[i] - \hat {\boldsymbol r}_{SAS}[i] \|^2, \\
     & ~s.t.~ \rm{Tr}({\boldsymbol \Phi}_{{eq_k}_{SAS}}[i]{\boldsymbol \Phi}^\emph{H}_{{eq_k}_{SAS}}[i])\leq \rm{P_R},
\end{aligned}
\end{equation}
where $\hat {\boldsymbol r}_{SAS}[i] = \sum_{k=1}^{n_rN}{\boldsymbol \Delta}_k{\boldsymbol G}_{{eq_k}_{SAS}}[i]{\boldsymbol \Phi}_{{eq_k}_{SAS}}[i]\hat{\boldsymbol s}[i]$. After we obtain the most-likely transmitted symbol vector $\hat {\boldsymbol s}[i]$ from the source node by ML detection as mentioned in the previous section, the RLS optimization problem is written as
\begin{equation}\label{4.16}
\begin{aligned}
    & [\hat {\boldsymbol \Phi}_{{eq_k}_{SAS}}[i]]\\ & = \arg\min_{{\boldsymbol \Phi}_{{eq_k}_{SAS}}[i]} \sum_{n=1}^{i}\lambda^{i-n}\|{\boldsymbol r}_{SAS}[n] - \hat {\boldsymbol r}_{SAS}[i]\|^2.
\end{aligned}
\end{equation}
By expanding the right-hand side of (\ref{4.16}) and taking gradient with respect to $({\boldsymbol \Phi}^*_{{eq_k}_{SAS}}[i])$ and equaling the terms to zero, we obtain
\begin{equation}\label{4.17}
\begin{aligned}
    & {\boldsymbol \Phi}_{{eq_k}_{SAS}}[i] \\ & = (\sum_{n=1}^{i}\lambda^{i-n}{\boldsymbol \Delta}_k{\boldsymbol G}_{{eq_k}_{SAS}}[i]{\boldsymbol \Phi}_{{eq_k}_{SAS}}[i]{\boldsymbol s}[i]{\boldsymbol r}^\emph{H}_{k_{SAS}}[n]) \\ & ~~~~ (\sum_{i=1}^{n}\lambda^{i-n}{\boldsymbol G}_{{eq_k}_{SAS}}[i]{\boldsymbol s}[i]{\boldsymbol r}^\emph{H}_{k_{SAS}}[n])^{-1}.
\end{aligned}
\end{equation}
The power constraint is enforced by the normalization strategy after the optimization. We can define
\begin{equation}\label{4.18}
\begin{aligned}
    {\boldsymbol \Psi}_{SAS}[i]  & = \sum_{n=1}^{i}\lambda^{i-n}{\boldsymbol r}_{k_{SAS}}[n]{\boldsymbol r}^\emph{H}_{k_{SAS}}[n] \\ & = \lambda {\boldsymbol \Psi}_{SAS}[i-1]+{\boldsymbol r}_{k_{SAS}}[n]{\boldsymbol r}^\emph{H}_{k_{SAS}}[n],
\end{aligned}
\end{equation}
\begin{equation}\label{4.19}
\begin{aligned}
    {\boldsymbol Z}_{SAS}[i] & = \sum_{n=1}^{i}\lambda^{i-n}{\boldsymbol r}_{e_{SAS}}[n]{\boldsymbol r}^\emph{H}_{k_{SAS}}[n] \\ & =\lambda {\boldsymbol Z}_{SAS}[i-1]+{\boldsymbol r}_{e_{SAS}}[n]{\boldsymbol r}^\emph{H}_{k_{SAS}}[n],
\end{aligned}
\end{equation}
where ${\boldsymbol r}_{e_{SAS}}[n]={\boldsymbol \Delta}_k{\boldsymbol G}_{{eq_k}_{SAS}}[i]{\boldsymbol \Phi}_{{eq_k}_{SAS}}[i]{\boldsymbol s}[i]$ and ${\boldsymbol r}_{k_{SAS}}[n]={\boldsymbol G}_{{eq_k}_{SAS}}[i]{\boldsymbol s}[i]$. Thus, we can rewrite (\ref{4.17}) as
\begin{equation}\label{4.20}
    {\boldsymbol \Phi}_{{eq_k}_{SAS}}[i] = {\boldsymbol Z}_{SAS}[i]{\boldsymbol \Psi}^{-1}_{SAS}[i].
\end{equation}
By employing the matrix inversion lemma in \cite{Tylavsky}, we can obtain
\begin{equation}\label{4.21}
\begin{aligned}
    {\boldsymbol \Psi}^{-1}_{SAS}[i] = & \lambda^{-1}{\boldsymbol \Psi}^{-1}_{SAS}[i-1]\\ & -\lambda^{-1}{\boldsymbol k}_{SAS}[i]{\boldsymbol r}^\emph{H}_{k_{SAS}}[i]{\boldsymbol \Psi}^{-1}_{SAS}[i-1],
\end{aligned}
\end{equation}
where ${\boldsymbol k}_{SAS}[i]=(\lambda^{-1}{\boldsymbol \Psi}^{-1}_{SAS}[i-1]{\boldsymbol r}_{k_{SAS}}[i])/(1+\lambda^{-1}{\boldsymbol r}^\emph{H}_{k_{SAS}}[i]{\boldsymbol \Psi}^{-1}_{SAS}[i-1]{\boldsymbol r}_{k_{SAS}}[i])$. We define ${\boldsymbol P}_{SAS}[i]={\boldsymbol \Psi}^{-1}_{SAS}[i]$ and by substituting (\ref{4.19}) and (\ref{4.21}) into (\ref{4.20}), the expression of the code matrix is given by
\begin{equation}\label{4.22}
\begin{aligned}
    & {\boldsymbol \Phi}_{{eq_k}_{SAS}}[i]\\
    &= \lambda{\boldsymbol Z}_{SAS}[i-1]{\boldsymbol P}_{SAS}[i]+{\boldsymbol r}_{e_{SAS}}[i]{\boldsymbol r}^\emph{H}_{k_{SAS}}[i]{\boldsymbol P}_{SAS}[i]\\
    &= {\boldsymbol Z}_{SAS}[i-1]{\boldsymbol P}_{SAS}[i-1]+{\boldsymbol Z}_{SAS}[i-1]{\boldsymbol k}_{SAS}[i]\\
    & ~~~{\boldsymbol r}^\emph{H}_{k_{SAS}}[i]{\boldsymbol P}_{SAS}[i-1]+{\boldsymbol r}_{e_{SAS}}[i]{\boldsymbol r}^\emph{H}_{k_{SAS}}[i]{\boldsymbol P}_{SAS}[i]\\
    &= {\boldsymbol \Phi}_{{eq_k}_{SAS}}[i-1]+\lambda^{-1}({\boldsymbol r}_{e_{SAS}}[i]-{\boldsymbol Z}_{SAS}[i-1]\\
    & ~~~ {\boldsymbol k}_{SAS}[i]){\boldsymbol r}^\emph{H}_{k_{SAS}}[i]{\boldsymbol P}_{SAS}[i-1].
\end{aligned}
\end{equation}
By using the algorithm in Table I we can obtain the optimal code matrices, and the only differences are the initialized matrices ${\boldsymbol P}[0]=\delta^{-1}{\boldsymbol I}_{N \times N}$, and ${\boldsymbol Z}[0] = {\boldsymbol I}_{N \times N}$ for SAS.

\section{Simulations}

The simulation results are provided in this section to assess the
proposed scheme and algorithms. The cooperative MIMO system
considered employs AF and DF protocol with the Alamouti STBC scheme
in \cite{RC De Lamare} using QPSK modulation in a quasi-static block
fading channel with AWGN. The effect of the direct link is also
considered. It is possible to employ the different DSTCs with a
simple modification. The cooperative MAS equipped with $n_r=1,2$
relay nodes and $N=2$ antennas at each node, while the cooperative
SAS employed $N=1,2$ antennas at the source node and the destination
node and $n_r=1,2$ relay nodes with single-antenna. In the
simulations, we set the symbol power $\sigma^2_s$ as equal to 1, and
the power of the adjustable code matrix in the DT-ACMO algorithm is
normalized. The $SNR$ in the simulations is the received $SNR$ which
is calculated by $SNR = \frac{\parallel{\boldsymbol
D}_D[i]\parallel^2_F}{\parallel{\boldsymbol n}_D[i]{\boldsymbol
n}^\emph{H}_D[i]\parallel^2_F}$.

The BER performance of the D-Alamouti, the R-Alamouti in
\cite{Birsen Sirkeci-Mergen} in the cooperative MAS and SAS without
delay are shown in Fig. 2. The solid strings are the BER curves
without the direct link (DL) and the dashed ones are that with the
DL. The SAS employing single antenna at each node and the AF
strategy with and without the DL have the worst BER performance
compared with the SAS and MAS employing multiple antennas and relay
nodes. The cooperation of the DL helps SAS to achieve a higher
diversity gain and lower the BER. In SAS employing $N=2$ antennas at
the source node and the destination node with $n_r=2$ single-antenna
relay nodes, and MAS employing the same number of the antennas at
the source node and the destination node with $n_r=1$ relay node
using $N=2$ antennas, the superposition of the BER curves can be
observed. With the cooperation of the DL, a higher order of the
diversity can be achieved. The simulation results lead to the
conclusion that with the same number of antennas at the source node
and the destination node in SAS and MAS, if the number of the relay
nodes with single antenna is the same as the number of multi-antenna
relay nodes, the same diversity order and BER performance can be
achieved when the synchronization is perfect at the relay node.

\begin{figure}
\begin{center}
\def\epsfsize#1#2{1\columnwidth}
\epsfbox{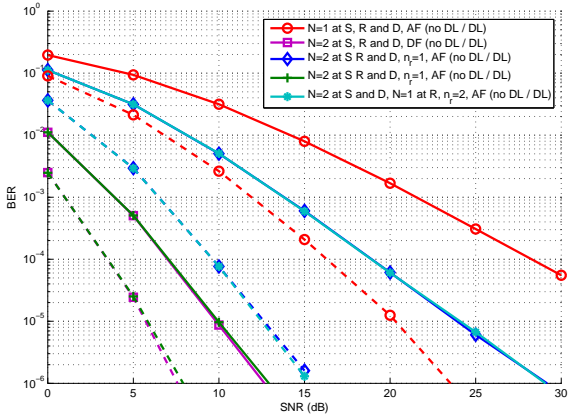} \caption{BER vs. $SNR$ Performance for the SAS
and MAS employing  the Alamouti Schemes without the Delay
Profile}\label{1}
\end{center}
\end{figure}

In Fig. 3, BER curves of different Alamouti coding schemes without
the DL using an ML detector are compared in SAS with $N=2$ relay
nodes. The D-Alamouti STBC is employed among the single-antenna
relay nodes. The different delay profiles are considered to evaluate
the D-Alamouti STBC in SAS. As shown in Fig. 3, the R-Alamouti
scheme improves the performance by about $2$dB without the DL
compared to the D-Alamouti scheme when perfect synchronization is
considered. When the delay profile ${\boldsymbol \Delta}=[0,1]$ is
considered among the relay nodes, it is shown that the D-Alamouti
and R-Alamouti are not DT-STCs.

\begin{figure}
\begin{center}
\def\epsfsize#1#2{1\columnwidth}
\epsfbox{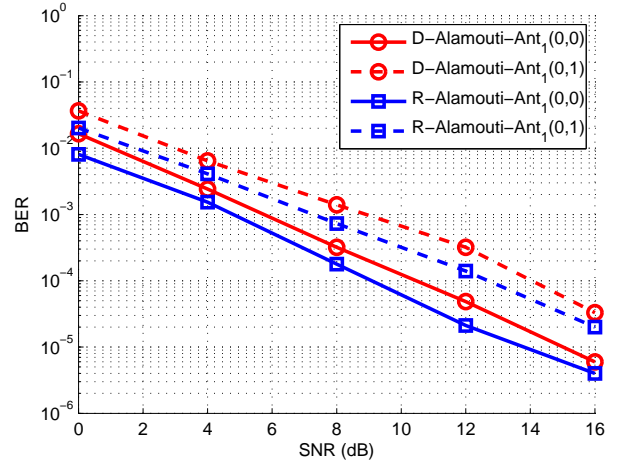}  \caption{BER vs. $SNR$ Performance for
SAS}\label{2} \vspace{-2em}
\end{center}
\end{figure}

The proposed DT-ACMO algorithm with an ML receiver is compared with
different Alamouti schemes in MAS in Fig. 4. It is worth to mention
that the delay profiles in the simulations are different. In the
proposed algorithm the DF strategy is employed which requires
re-encoding at the relay nodes, and the full Alamouti scheme is
obtained at each of the $n_r=2$ relay nodes. The results illustrate
that without the DL, by making use of the randomized encoding
technique in \cite{Birsen Sirkeci-Mergen}, a $2$dB BER performance
improvement can be achieved compared to the D-Alamouti scheme with
and without the delay. It is also worth to mention that the DT-ACMO
algorithm can be employed in the MAS to achieve a better BER
performance as both of algorithms employ DSTCs and can perform the
optimization at the destination node. With the consideration of the
delay profiles, the results indicate that the diversity order will
not reduced, and using the DT-ACMO RLS algorithm an improved
performance is achieved with $3$dB of gain as compared to employing
the RSTC algorithm in \cite{Birsen Sirkeci-Mergen} and $5$dB of gain
as compared to employing the traditional DSTBC algorithm. The
simulation results illustrate that for MAS with different DSTC
schemes when the same delay profiles are considered, a $1$dB
performance improvement is obtained. It is because when the full
DSTC schemes can be obtained at the relay nodes employed in MAS, at
the destination node a delayed symbol matrix with increased
dimension is received which means the observation window is bigger
than that with the non-delayed MAS.

\begin{figure}
\begin{center}
\def\epsfsize#1#2{1\columnwidth}
\epsfbox{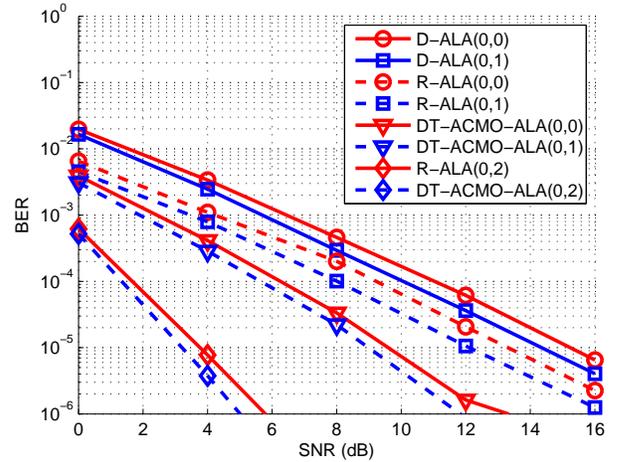}  \caption{BER vs. $SNR$ Performance for DT-ACMO
RLS Algorithm for MAS}\label{2}

\end{center}
\end{figure}

\section{Conclusion}

We have proposed a delay-tolerant adjustable code matrix
optimization (DT-ACMO) algorithm for the cooperative MIMO system
using an ML receiver at the destination node to mitigate the effect
of the delay associated with the DSTCs from relay nodes. Two types
of cooperative systems have been analyzed by comparing the BER of
the delayed DSTCs. In order to achieve a better BER performance and
eliminate the computation complexity at the destination node, we
have proposed the DT-ACMO RLS algorithm. The simulation results
illustrate the performance of the two different cooperative systems
with the same number of transmit and receive antennas, and the
advantage of the proposed DT-ACMO and DT-ACMORO algorithms by
comparing them with the cooperative network employing the
traditional delayed DSTC scheme and the delayed RSTC scheme. The
proposed algorithms can be used with different DSTC schemes and can
also be extended to cooperative systems with any number of antennas.

\appendices

\bibliographystyle{IEEEtran}

\end{document}